\documentclass[fleqn,usenatbib]{mnras}

\usepackage{newtxtext,newtxmath}
\usepackage{array}
\usepackage[acronym,nonumberlist]{glossaries}
\usepackage{xcolor}
\usepackage{bm}

\newacronym{pa}{PA}{Position Angle}
\newacronym{rms}{RMS}{Root Mean Square}
\newacronym{pd}{PD}{Phase Diagram}

\usepackage[T1]{fontenc}

\DeclareRobustCommand{\VAN}[3]{#2}
\let\VANthebibliography\thebibliography
\def\thebibliography{\DeclareRobustCommand{\VAN}[3]{##3}\VANthebibliography}

\usepackage{graphicx}	

\usepackage{amsmath} 

\title[Rotation Period of C/2006 P1 (McNaught)]{Rotation Period of C/2006 P1 (McNaught) 
Through Morphological Analysis with Narrowband Imaging}

\author[V. Okoth et al.]{
Vincent Okoth,$^{1}$\thanks{E-mail: vincent.okoth.ac.uk (VO)}
Cyrielle Opitom$^{1}$
Colin Snodgrass $^{1}$
Brian Murphy $^{1}$
James E. Robinson $^{1}$
\\
$^{1}$Institute for Astronomy, University of Edinburgh, Royal Observatory, Edinburgh, EH9 3HJ, UK\\
}

\date{Accepted 2026 February 17. Received 2026 February 13; in original form 2025 December 5}

\pubyear{\the\year{}}

\begin{document}
\label{firstpage}
\pagerange{\pageref{firstpage}--\pageref{lastpage}}
\maketitle

\begin{abstract}
This study presents findings from narrowband imaging of comet C/2006 P1 (McNaught) using the 3.6-meter New Technology Telescope (NTT) at La Silla, Chile. Observations commenced on January 27, 2007, 15 days after perihelion, and continued until February 4, with additional sessions from February 25 to 28. Imaging was conducted using the ESO Multi-Mode Instrument (EMMI) in both broadband (B, V, R) and six comet-specific narrowband filters (CN, C$_3$, C$_2$, NH$_2$, blue and red continuum). Various image processing techniques were employed to enhance structural features, including azimuthal mean/median division and subtraction, azimuthal renormalization, and division by a 1/$\rho$ profile as well as Larson-Sekanina technique. These enhancements revealed dynamic coma structures, with jets transitioning from spiral patterns to linear or fan-like shapes over time. The consistency of morphological patterns across different processing methods validated their authenticity. The periodic recurrence and temporal evolution of CN coma features in narrowband images indicate a nucleus rotation period of 11.8 ± 0.5 h, consistent with stable active regions and rotationally modulated outgassing near perihelion.

\end{abstract}




\section{Introduction}
Comet C/2006 P1 (McNaught), hereafter referred to as C/2006 P1, a long-period comet, captured global attention as the brightest comet observed in over four decades, becoming the most luminous since C/1965 S1 (Ikeya-Seki), and earning the title of the “Great Comet of 2007” \citep{2007MNSSA..66...61.}. Discovered on 7th August, 2006, by Robert H. McNaught at the Siding Spring Observatory, it displayed remarkable brightness and a visually stunning striated tail exceeding 35$^{\circ}$, during its perihelion passage \citep{2010KPCB...26..322K}. 

The comet follows a highly eccentric trajectory with an orbital period exceeding 92,000 years, classifying it as a long period, dynamically new comet \citep{2007IAUC.8801....2M}. It reached perihelion on 12, January 2007, at a heliocentric distance of just 0.17 au, closer to the Sun than Mercury, and attained a peak brightness of magnitude –5.5. The combination of close solar approach, volatile-rich composition, and intense activity in mid-January 2007 produced a dense coma and a striking dust tail, providing a historic visual spectacle. Its exceptional observational characteristics, combined with its pristine nature as a dynamically 'new' comet originating from the Oort Cloud, make it an important target for investigating the physical and chemical properties of cometary nuclei and comae \citep{2016AJ....152..220S}.

Studies of C/2006 P1 have highlighted its intense dust and gas production rates, intricate coma morphology, and dynamically evolving tails \citep{2019Icar..319..540P,2010KPCB...26..322K,2007ApJ...667.1262N,2008A&A...482..293L,2007ApJ...661L..93F,2007DPS....39.5311D,2007IAUC.8801....2M}. High-resolution imaging by \citealt{2007CBET..832....1S} revealed dynamic jets and spirals arising from localized nucleus activity driven by sublimation of ices. Spectroscopic and narrowband studies identified a rich volatile inventory \citep{2007DPS....39.5311D,2007IAUC.8801....2M}, with emission bands from CN, C$_2$, C$_3$, NH$_2$, and dust continuum indicating substantial sublimation-driven activity during its near-Sun apparition. Space-based solar observatories, including the STEREO Heliospheric Imager, detected an arch-like tail composed of neutral Fe atoms, the first such observation \citep{2007ApJ...661L..93F}

Lessons from ESA’s Rosetta mission to 67P/Churyumov–Gerasimenko (hereafter referred to as 67P) have shown that cometary activity often arises from discrete sources influenced by topographic shading, diurnal cycles, and internal layering. \cite{2019SSRv..215...30V} reported that many jets on 67P originated from fractures and pits where volatiles were exposed to direct insolation. These features depended strongly on spin-axis orientation and seasonal effects. Although such detailed surface data are unavailable for C/2006 P1, similar principles can guide the interpretation of its coma morphology from Earth-based observations.

The connection between coma features and nucleus rotation has been demonstrated in several comets. Expanding spirals and arcs in narrowband CN and C$_2$ images are often modeled as periodic outgassing from fixed active regions on rotating nuclei \citep{2004come.book..449S}. In comets such as 1P/Halley \citep{1984AJ.....89.1408S}, C/1995 O1 (Hale-Bopp) \citep{1997EM&P...77..189S,2000ApJ...529L.107S}, and 103P/Hartley 2 \citep{2011AJ....141..183K}, such modeling has yielded constraints on rotation rates, orientation of the spin axis, active-region locations, and seasonal variability.

Narrowband filter imaging remains a powerful ground-based technique for cometary science. Standardised filter sets isolate molecular emissions (e.g., CN, C$_2$, C$_3$, NH$_2$, OH) and continuum features, separating gas from dust contributions \citep{2004come.book..449S,1990Icar...88..228O,2000Icar..147..180F}. Multi-decade surveys have used these filters to measure composition, production rates, and morphological changes driven by rotation or outbursts. CN, a daughter species of HCN \citep{1980ApJ...237..633C, 2005P&SS...53.1243F}, is a particularly sensitive tracer of active regions due to its localized release, optical prominence compared other gas species and rotational variability against dust. Time-series CN imaging has mapped jet curvature, coma morphology evolution and rotation period in comets such as 21P/Giacobini–Zinner \citep{2023PSJ.....4...28G}, 103P/Hartley 2 \citep{2011AJ....141..183K,2012A&A...543A..32W}, 45P/Honda–Mrkos–Pajdušáková \citep{2022PSJ.....3...15S}, and 46P/Wirtanen \citep{2021PSJ.....2..104K,2021PSJ.....2....7F}.

Despite its historic brightness, C/2006 P1 remains less studied than other Great Comets such as  C/1995 O1 (Hale-Bopp), hereafter referred to as Hale-Bopp, or C/2020 F3 (NEOWISE), referred to as F3 for the remainder of this work. Unlike Hale-Bopp, where extensive imaging campaigns were carried out from multiple hemispheres, C/2006 P1’s southern trajectory, its close proximity to the sun around perhelion, and rapid brightness evolution limited continuous observational coverage. Although narrowband imaging during perihelion revealed a coma rich in CN, C$_2$, C$_3$, and other volatiles with marked spatial asymmetries \citep{2007CBET..832....1S}, these morphological structures have not been examined in detail through time-series analyses or linked to rotational modulation. Existing studies have largely prioritized overall production rates over detailed coma morphology, leaving a substantial opportunity for retrospective, time-series  and high-resolution morphological analysis.

This study aims to address some of these questions by analyzing the coma morphology and rotational state of C/2006 P1 through analysis of time-resolved narrowband imaging data. By examining the evolution of structural features such as linear jets, fans, arcs and spirals in CN, this work investigates temporal and spatial variations in the coma, and further places new constraints on the nucleus’s rotation period.

\section{Observation and Data Reduction}
Observations of C/2006 P1 (McNaught) were obtained during its perihelion passage in January–February 2007 from the European Southern Observatory’s (ESO) La Silla facility in Chile. Data were collected with the 3.6 m New Technology Telescope (NTT) equipped with the ESO Multi-Mode Instrument (EMMI), a versatile imager and spectrograph capable of direct imaging, low, medium, and high-resolution spectroscopy, and echelle spectroscopy (this work focuses on narrowband imaging). Mounted at the Nasmyth B focus, EMMI covered 300–1000 nm, divided between a blue arm (300–500 nm) and a red arm (400–1000 nm), with rapid mode switching capability. EMMI offered spatial resolutions down to 0.13\arcsec for high-resolution imaging, 0.1665\arcsec for red wide-field imaging, and 0.37\arcsec for blue wide-field imaging, suitable for both point sources and extended targets. For these observations, $2\times2$ on-chip binning was employed in the red arm, yielding a pixel scale of 0.33\arcsec per binned pixel. Blue arm observations were left unbinned.

Observations were conducted in two observing runs. The first campaign, from 27 January to 4 February, 2007, carried out approximately two weeks after perihelion, pushed the NTT to its operational limits, with the comet observed at extremely low elevation angles and reaching a minimum altitude of only 10$^{\circ}$. A second campaign from February 25 to 28 provided additional temporal coverage. Owing to the comet’s solar elongation, all imaging was conducted during both dusk and dawn twilight in each of the two observing runs, with usable windows of approximately 30 minutes per night.

The dataset comprises 210 exposures obtained through a combination of broadband and narrowband filters. The broadband filters used (B, V, R) were selected to capture the continuum emission of the comet, while six specialized narrowband cometary filters, the ESA-produced filters set for Rosetta mission were employed to isolate specific gas and dust emission features \citep{1996P&SS...44..619S}. These narrowband filters targeted the emission bands of CN (386 nm), C$_3$ (405 nm), C$_2$ (510 nm), and NH$_2$ (662 nm), as well as the blue and red dust continuum at 441 nm and 683 nm, respectively. In this work, we focused on the blue arm filters, CN, C$_3$ and blue continuum, shown in table \ref{tab:obs}.

Exposure times ranged from 5 to 90 seconds, depending on filter throughput, signal-to-noise ratio (S/N), comet brightness, and observing conditions. Non-sidereal tracking was used to track the comet’s apparent motion and keep it centered in the field. Calibration frames, including bias and twilight flat-fields were acquired during each session. Data reduction followed standard procedures: bias subtraction to remove detector electronic noise, flat-field correction to account for pixel-to-pixel sensitivity variations, and image cropping to remove overscan regions. No flux calibration was performed, as the analysis focused exclusively on coma morphology.

\begin{figure}
	\includegraphics[width=\columnwidth]{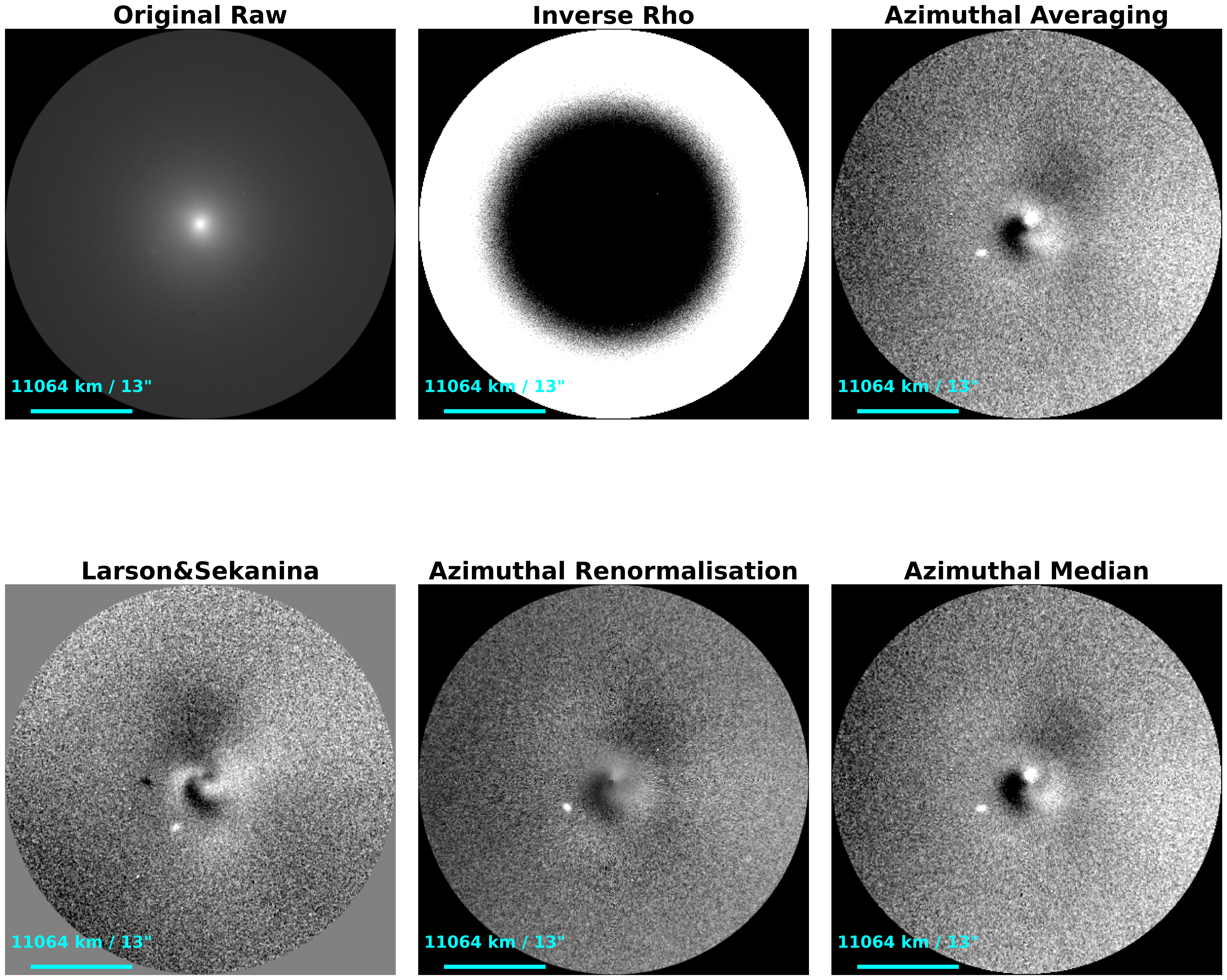}
    \caption{Representative CN image from 03/02/2007 (top left, $\sim$22 000 km/26 arcsec radius) alongside five enhancement techniques: 1/$\rho$ division, azimuthal average division, Larson–Sekanina, azimuthal median and azimuthal renormalisation respectively. Both the azimuthal mean and median methods are powerful for enhancing the structures but the azimuthal mean is more susceptible to noise and outliers, whereas the azimuthal median gives cleaner, more stable features. Azimuthal renormalisation equalises brightness variations, improving subtle contrast. The 1/$\rho$ division corrects the radial fall-off while Larson–Sekanina(LS) filter isolates curved and rotationally modulated features but produces known artefacts. A comprehensive comparison of these methods shows that persistent features are real, with the azimuthal median adopted for further analysis because of its stability and minimal artefacts and LS for its sensitivity to rotational features.}
    \label{fig:techniques}
\end{figure}

The quality of the dataset was generally good, though the low observing altitude, and high and varying sky background imposed limitations with limited temporal coverage as well as varying signal to noise(S/N) ratios posing a challenge for temporal and rotational analysis. Seeing ranged from 0.5\arcsec to 1.5\arcsec. Frames affected by cloud cover, saturation or poor tracking were discarded. Also, some nights were lost to bad weather as no CN or science images were obtained. A summary of the observations, including the number of frames per filter, observational and dynamical properties like heliocentric and geocentric distances as well as sun position angle and phase angles during the observing period is given in table~\ref{tab:obs}.

\begin{table*}
\centering
\caption{\textbf{Nightly Observations with Statistics for the Filters Used in our Analysis}}
\label{tab:obs}
\begin{tabular}{|>{\centering\arraybackslash}m{1.5cm}|
                >{\centering\arraybackslash}m{1.5cm}|
                >{\centering\arraybackslash}m{2.5cm}|
                >{\centering\arraybackslash}m{1.5cm}|
                >{\centering\arraybackslash}m{1.5cm}|
                >{\centering\arraybackslash}m{1.5cm}|
                >{\centering\arraybackslash}m{1.5cm}|}
\hline
\textbf{Date} & \textbf{Total Observations} & \textbf{Filters Used} & \textbf{Heliocentric Distance (au)} & \textbf{Geocentric Distance (au)} & \textbf{Phase Angle (deg)} & \textbf{Anti-Solar P.A. (deg))} \\
\hline
2007-01-29 & 5 & CN: 2, Bc: 2, C$_3$: 1 & 0.58 & 1.07 & 65.25 & 171.52 \\
\hline
2007-01-31 & 31 & CN: 12, Bc: 11, C$_3$: 8 & 0.63 & 1.12 & 61.31 & 175.44 \\
\hline
2007-02-02 & 8 & CN: 6, Bc: 2 & 0.70 & 1.17 & 57.46 & 175.51 \\
\hline
2007-02-03 & 24 & CN: 10, Bc: 10, C$_3$: 4 & 0.71 & 1.18 & 56.40 & 176.08 \\
\hline
2007-02-04 & 7 & CN: 4, Bc: 3 & 0.74 & 1.20 & 54.97 & 180.38 \\
\hline
2007-02-26 & 16 & C$_3$: 8, CN: 8 & 1.23 & 1.54 & 40.02 & 186.78 \\
\hline
2007-02-27 & 12 & C$_3$: 6, CN: 6 & 1.25 & 1.55 & 39.57 & 187.37 \\
\hline
2007-02-28 & 10 & CN: 5, C$_3$: 5 & 1.27 & 1.56 & 39.22 & 187.78 \\
\hline
\end{tabular}
\caption{Summary of nightly observations of comet C/2006 P1 (McNaught) using multiple filters i.e., CN, C$_3$ and blue continuum (Bc in the table). The table shows heliocentric distance, geocentric distance, phase angle, and position angle of the Sun. Observations span from January 27 to February 28, 2007, with some nights missing as science images were not obtained due to bad weather. A total of \textbf{52 CN filter frames} were acquired across these observation nights.}
\end{table*}

\section{Results and Analysis}

\subsection{Coma Morphology}

\begin{figure}
    \centering
    \includegraphics[width= \columnwidth]{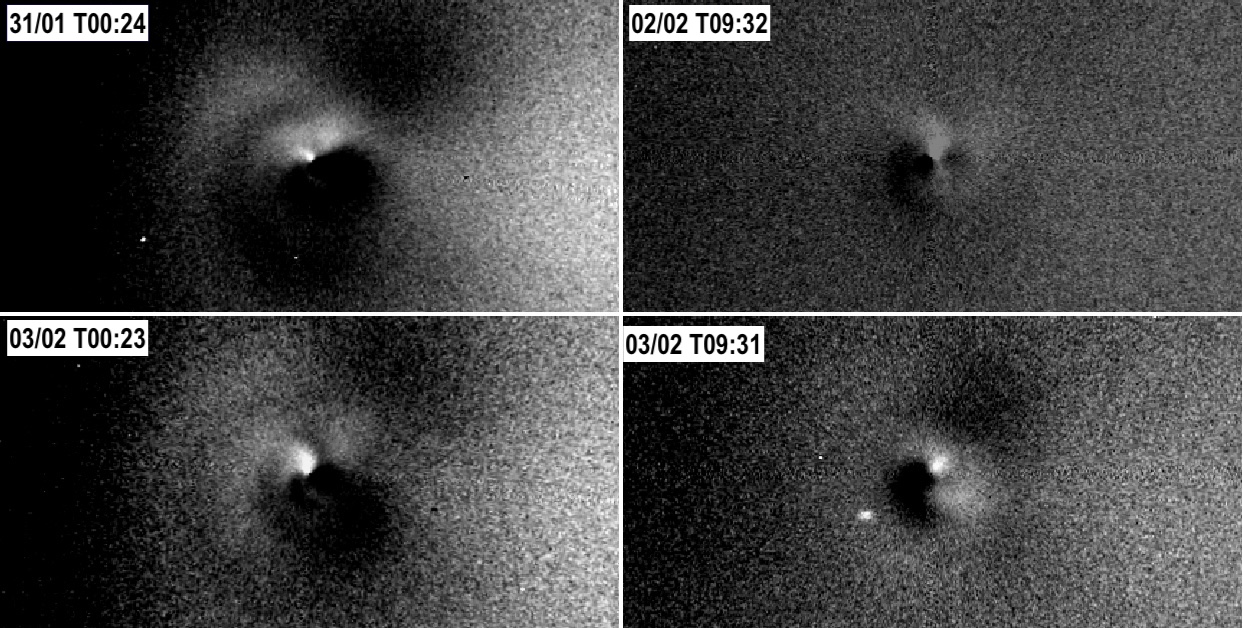}
    \caption{Representative images of CN morphology enhanced with azimuthal median removal technique during the first observing run arranged chronologically, from 31 Jan (Top left) to 03 Feb, 2007 (bottom right). The images showed linear jets (one or two) at the nucleus, spirals and arcs possibly originating from the nucleus dominating throughout the observing period }
    \label{fig:Naught1}
    
    \vspace{1cm} 
    
    \includegraphics[width=\columnwidth]{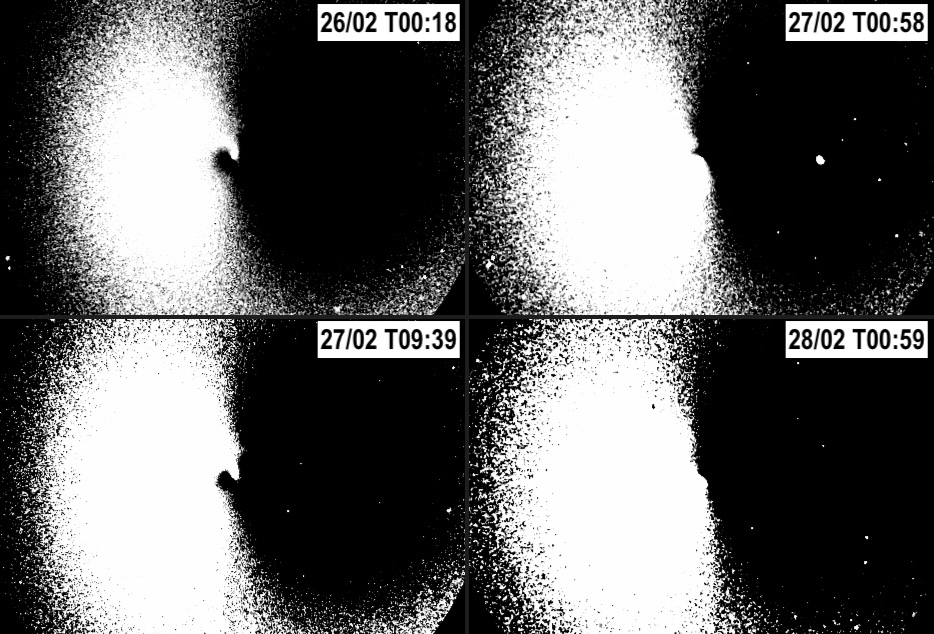}
    \caption{Chronological representation of CN images enhanced with azimuthal median removal technique, observed in the second run, 26-28 Feb 2007. The morphology is dominated by wide fan jets and overall asymmetric features}
    \label{fig:Naught2}
\end{figure}

\begin{figure*}
	\includegraphics[width=0.85\textwidth]{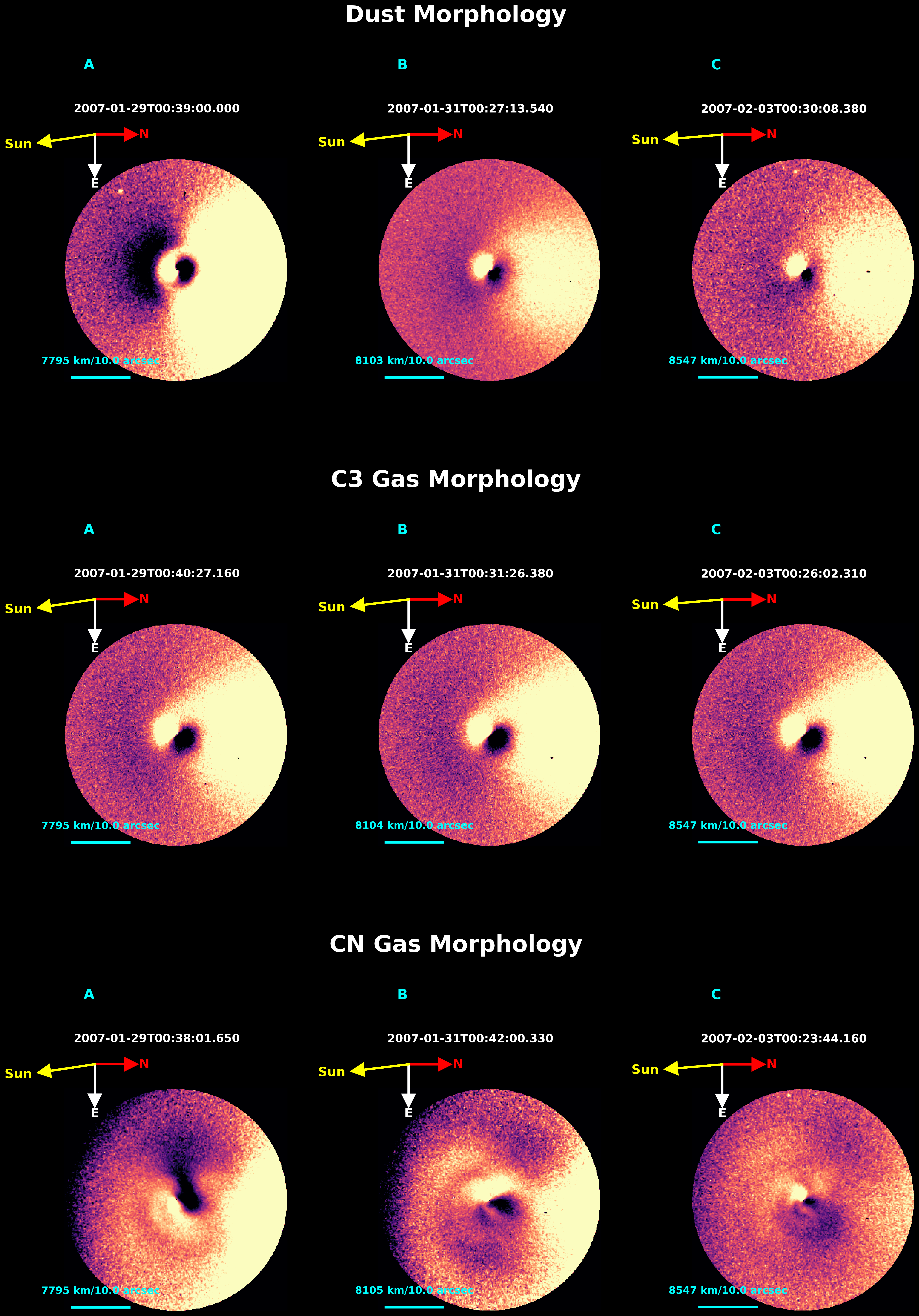}
    \caption{Images representing dust coma morphology (top panels), C$_3$ morphology (middle panels) and CN morphology (bottom panels) taken at nearly similar times(from 29 January to 3, 2007). The dust coma shows broad, anti-sunward, fan-like structures and the C$_3$ images exhibit comparably similar pattern but smoother, probably dominated by dust, possibly because the wide C$_3$ bandpass allows significant dust contamination, producing stable, dust-like morphologies showing no signs for rotational modulation. The observed C$_3$ structures might also reflect its weak, less optically dominant emission hence dominated by the slower, more uniform dust outflow that masks the short-term variability. In contrast, CN images reveal sharper, diverse, evolving jets possibly tracing nucleus rotation with less or minimal dust contamination. The orientation is such that east is to the bottom (white arrow) and north to the right (red arrow) with the sun direction shown in yellow arrow, and the time is in universal time(UT).}
    \label{fig:FilterComparison}
\end{figure*}

\begin{figure*}
	\includegraphics[width=0.9\textwidth, height=0.9\textheight, keepaspectratio=true]
    {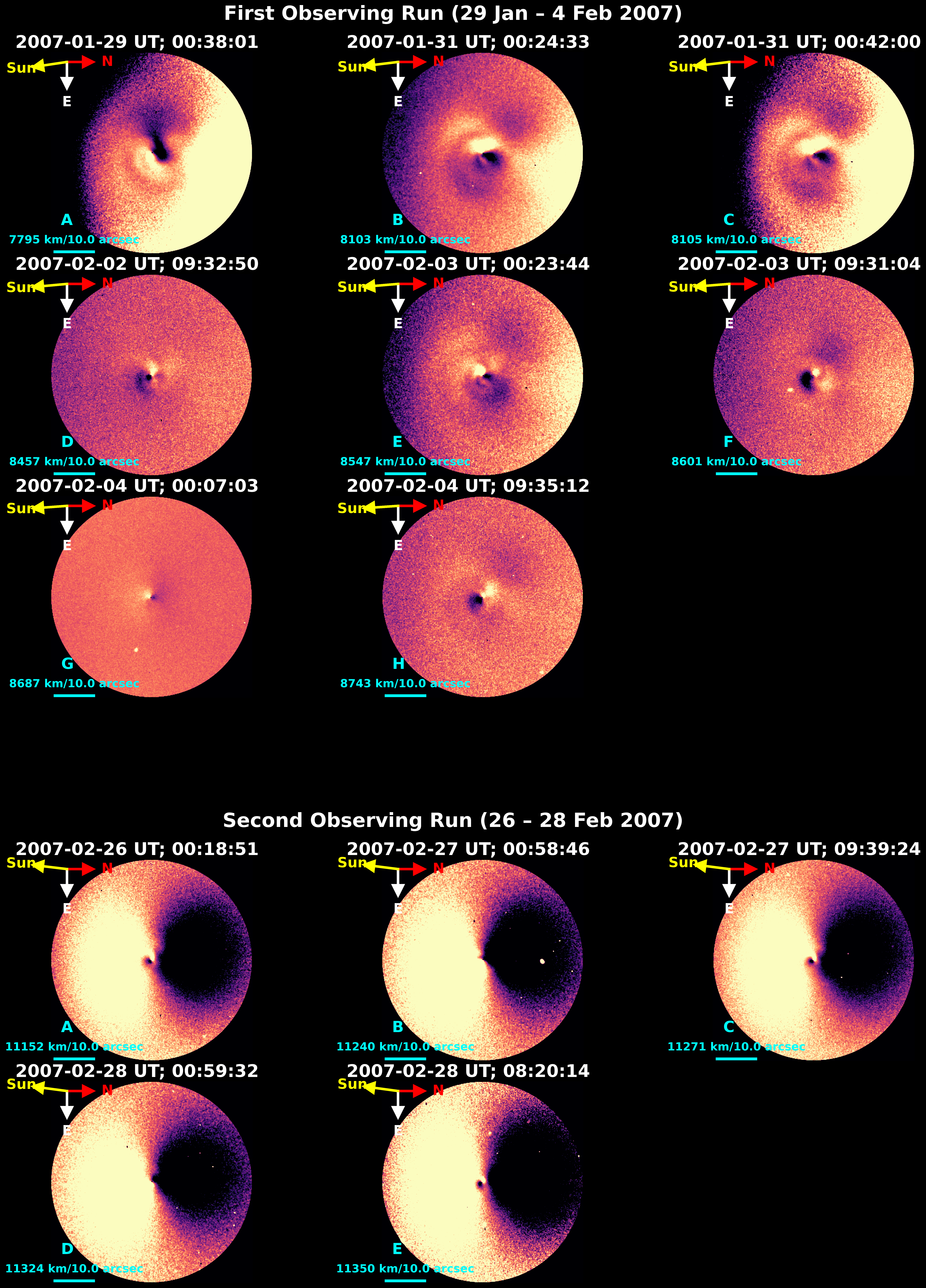}
    \caption{Time-series of enhanced CN images from 29 January–28 February 2007 showing the temporal evolution of anisotropic coma structures in C/2006 P1. The morphology evolves from arcs and spiral-like shells to narrow, rotating jets and broad fan-shaped features. Recurrent appearance of the same jets at different epochs enabled tracking of their position angles. The repetition of identical structures, shifting by tens to hundreds of degrees over successive nights, suggests rotational modulation. Notably, near similar jet orientations and morphology were observed on  31 Jan, 00:24(UT) and 3 Feb, 00:23 UT, 26 Feb, 00:18(UT), 27 Feb, 09:39(UT) and 28 Feb, 08:20(UT), and  27 Feb, 00:58(UT) and 28 Feb, 00:59(UT), with comparable structures reappearing after $\sim22.5$, $\sim33.5$ or $\sim72.0$ hours, suggesting a repeating pattern with a factor near 12 hours, though the period could not be estimated from this approach. The orientation is such that east is to the bottom (white arrow) and north to the right (red arrow) with the sun direction shown in yellow arrow and the time is in UT.}
    \label{fig:CN_morphology}
\end{figure*}

The coma morphology of C/2006 P1 was investigated through the application of image enhancement techniques designed to reveal subtle features, such as jets, spirals, and fans often obscured by the comet’s bright and diffuse appearance \citep{2014Icar..239..168S,2006DPS....38.2913S}. These procedures enabled the identification of evolving morphological structures across different epochs and filters.

The first critical step was accurate image centering. Using 2D Gaussian centroiding, the optocenter was identified with 0.01 pixel accuracy as the center of light for each frame. To further refine this alignment, recentroiding was conducted four times on each image, progressively using smaller aperture from 137 to 11  pixel sizes each time on the resultant image to reduce systematic offsets from the algorithm. The images were then re-centered to the optocenter, ensuring consistency across the dataset. They were then aligned such that North is right and East is bottom, allowing all \glspl{pa} to be measured consistently with respect to north in clockwise direction. This high degree of precision was essential for the success of subsequent enhancement steps. Misalignments at this stage could have introduced distortions, compromising the visibility and reliability of morphological structures.

Following this data preprocessing, multiple feature enhancement techniques were applied to the centered images to emphasize the comet's azimuthal and radial structures. These methods were implemented using a combination of the Cometary Coma Image Enhancement Facility \citep{2015P&SS..118..181M} and customized Python algorithm optimized for jet structure enhancement \citep{2025P&SS..26806178M}. Each technique was designed to address specific challenges associated with the analysis of cometary features, with complementary strengths and limitations.

Azimuthal mean and median subtraction or division techniques calculate the average or median brightness at each radial distance and then subtract or divide the original values accordingly to improve the visibility of azimuthal features. The mean method emphasized such azimuthal variations, but could be affected by outliers from background stars or image noise etc, while the median method proved more robust against noise. These approaches provided high sensitivity to features such as jets and arcs, though they required precise centering to avoid the introduction of artefacts. Division by a $1/\rho$ radial profile was also employed to adjust for the radial brightness fall-off due to the expected profile for a steady state comet, though showed low sensitivity to azimuthal variations. Azimuthal renormalization was tested as another approach to standardize brightness variations across the azimuthal profile. Additionally, we used the Larson-Sekanina rotational filtering technique for further comparison \citep{1984AJ.....89..571L}. See \citealt{2014Icar..239..168S} for details on these and more cometary image processing methods.

Each enhancement technique was carefully evaluated for its ability to reveal morphological features consistently across the dataset. By comparing the results of different methods as shown in figure \ref{fig:techniques}, the reliability of observed structures was confirmed, as fairly consistent features were observed irrespective of the technique used. This comparative approach ensured that the enhancements were not artefacts of any specific method, thereby improving confidence in the analysis. Also, the combined use of these techniques ensured a comprehensive assessment of the coma morphology. Due to its known effectiveness in suppressing noise and introducing minimal artifacts \citep{2015A&A...574A..38O}, the azimuthal median method was preferred and ultimately chosen for further analysis. For completeness of analysis, we also selected images enhanced by the Larson–Sekanina method, given its strong sensitivity to sharp, fine-scale, spiral and rotationally modulated features, although we note that this technique is known to introduce characteristic artefacts and is sensitive to rotation and shift parameters.

The application of these enhancement techniques revealed a range of dynamic morphological features in the comet’s coma. Enhanced CN-filter images, in particular, displayed distinct linear jets, spirals and arcs during the first observing epoch (January 29 to February 4, 2007 as shown in figure shown in figure \ref{fig:Naught1}), evolving into fan jets and asymmetric structures by the second epoch (February 25 to 28, 2007, shown in figure \ref{fig:Naught2}). Comparative analysis of filters showed that CN images consistently exhibited the most diverse morphological features, while C$_3$ and blue continuum filters highlighted broader fan structures (see figure \ref{fig:FilterComparison} and next section for details).

\subsection{Temporal Evolution of the Coma Morphology}
\subsubsection{Narrowband Morphology} 
Narrowband imaging of C/2006 P1 revealed clear differences in coma morphology between CN and other gas species. Unlike CN, C$_3$ and blue continuum images instead display broad, diffuse, and largely stable structures, probably showing dust structures following the anti-solar direction, with no evidence of rotational modulation as shown in the top and middle panels in \ref{fig:FilterComparison} from A through C in each panel, with bottom panels showing CN gas morphology for comparison. The morphology seen in the C$_3$ filter is likely dominated by dust contamination, with the continuum images similarly tracing dust-driven structures rather than gas emissions. In contrast, CN-filter images displayed diverse, high-contrast jet structures, including a narrow jet whose orientation evolved noticeably over time, possibly tracing a rotating, localized active source on the nucleus (see the bottom panels in figure \ref{fig:FilterComparison} from A through C and also figure \ref{fig:CN_morphology} from A through H, and A through E in the first and second runs respectively). CN also showed higher visibility and detectability of the jets compared to other gas species, with more robust and sharply defined features. Blue continuum images, used as a dust proxy, showed smooth, fan-like tails pointing away from the Sun, consistent with dust emission under radiation pressure. Comparison with CN images confirmed minimum detectable dust contamination in the CN filter, however, high dust contamination was detected in the other gas filters. 

The absence of rotational signatures in filters other than CN is likely due to both the limited temporal coverage on these filters and the physical properties of the emitting species; lighter, faster-moving gases such as CN can resolve and trace rotational modulation, whereas heavier or slower-moving dust expand more uniformly, producing broader, smoother structures that mask short-term variations. Moreover, the weak and less visible C$_3$ emission likely suffers from significant dust contamination, further obscuring any rotational features on them.

\subsubsection{CN Gas Morphology}
Figure~\ref{fig:CN_morphology} presents a time series of enhanced narrow-band CN images of C/2006 P1 obtained between 29 January and 28 February. Each panel from A through H, and A through E corresponds to a distinct epoch in the first and second observing runs respectively, revealing morphological changes in the inner coma. Assuming persistent yet evolving features throughout the dataset, suggest localized, anisotropic activity modulated by nucleus rotation, enabling tracking of jets and the assessment of the comet’s rotation period. The structures are interpreted as the 2D projection of the underlying 3D jet–coma geometry in the plane of the sky, mapping the active regions requires modeling the 3D outflow geometry.

In panel A in figure \ref{fig:CN_morphology}, on 29 January at 00:38 UTC, taking the collimated jet that was seen at a \gls{pa} of $\sim$127$^\circ$, extending $\sim$5,000 km ($\sim 6.5$ arcsec and beyond from the nucleus and assuming the same jet, by 31 January at 00:24 UTC (panel B), it had rotated to \gls{pa} $\sim$246$^\circ$, a total angular displacement of $\sim$119$^\circ$ or $\sim$241$^\circ$ in either direction. As shown in panel D, further rotation was observed by 2 February at 09:32 UTC, when the jet pointed westward (\gls{pa} $\sim$293$^\circ$), and by 3 February at 00:23 UTC (panel E), it had returned to \gls{pa} $\sim$242$^\circ$ (compare with panels B and C), suggesting the pattern repeats roughly every 24 hours, or an integer factor thereof.

Arc-like shell structures with radii of $\sim$3,000–15,000 km $\sim3.8-18.8$ arcsec appeared on nights 29 January, 31 January, and 3 February (see panels A, B, C and E). A corkscrew-shaped jet was observed on 3 February at 09:31 UTC (panel F) . On 4 February, more rapid evolution was evident. In panel G, at 00:07 UTC, the jet appeared subdued, possibly due to low signal to noise ratio due from bright sky background. However, by 09:35 UTC (panel H), the jet had rotated to \gls{pa} $\sim$311$^\circ$, marking a $\sim$297$^\circ$ shift within 9.5 hours, showing evidence of a relatively fast rotating nucleus.

By the second run (27–28 February), the jets had broadened into fan-like structures. Assuming the same jet, on 27 February at 00:58 UTC, it pointed sunward, rotating to \gls{pa} $\sim$225$^\circ$ by 09:39 UTC. On 28 February at 00:59 UTC, the structure appeared bifurcated, evolving into a diffuse, symmetric envelope by 08:20 UTC, possibly due to overlapping active sources or increased CN dispersion.

Across both runs, morphological recurrence and a $\sim$287$^\circ$ jet rotation over 9.5 hours suggest a nucleus rotation period greater $9.5$ hours, the minimum time difference between images, and around 12 hours (by extrapolating the rotation from angular displacement), consistent with a stable, relatively fast-rotating source region. While this approach does not allow us to determine the rotation period, it provides a qualitative picture of how the morphology evolved and how the jets shifted in \glspl{pa} over the course of the observations. It is also important to note that in general, high spatial resolution and good temporal cadence are vital for revealing recurring inner coma features. However, because the comet could only be observed at roughly the same local times each night, the range of accessible rotational phases was limited, particularly for periods that are integer factors of 24 hours. Moreover, in some observations, poor seeing and bright sky background obscured more subtle structures, and the short observing window (30 minutes before sunrise and sunset) due to comet's close proximity to the sun and long gaps between observations hindered accurate phase alignment and jet tracking, limiting reliable interpretation of temporal morphology changes.

\subsection{Rotation Period}
Building on the analysis of the temporal evolution of CN jet morphology, we used time-series imaging to estimate the rotation period of the nucleus. The multiple epoch time-series CN imaging revealed periodic jet structures whose morphology and  \glspl{pa} could be correlated across nights to estimate the rotation period. Pairs of images showing the same morphological configuration were identified to test for rotational recurrence. For example in figure \ref{fig:CN_morphology} in the second observing campaign, the CN morphology in panel A on 26 February at 00:18 closely resembled that on 27 February at 09:39 (panel C) and 28 February at 08:20, in panel E while the morphology on 27 February at 00:58 (panel B) was nearly identical to that in panel D on 28 February at 00:59, with almost matching orientation and curvature of the primary jet structures relative to the nucleus. The shortest elapsed time between two such images with almost similar morphologies was $\sim22.41\pm$0.90 h. Assuming an integer number of complete rotations between these observations, a range of trial periods was tested. The best match in rotational phase alignment across multiple image pairs was obtained with a period of 11.8$\pm$0.5 h(see next subsections for details).

Given the limitations of relying solely on morphological recurrence, particularly the sparse temporal coverage and potential projection effects, we applied two complementary methods to further constrain the period. These were: (i) computing \gls{rms} residuals of the ratio between image pairs and (ii) tracking the angular position and \glspl{pa} of distinct jet features (see the next sub-section for details). Similar or related techniques have been successfully applied to comets such as 1P/Halley, C/1995 O1 (Hale–Bopp), and C/2012 F6 (Lemmon) (e.g., \citealt{1998Icar..131..233S, 1998ApJ...501L.221L, 2015A&A...574A..38O}), where rotationally modulated coma structures were used to derive periods ranging from a few hours to over a day.

\subsubsection{Rotation Period by Root Mean Square}
This method involves first normalizing the images and then dividing sequential frame by one another to highlight temporal variations in coma morphology. The \gls{rms} of the resulting normalized divided images was then computed and plotted as a function of the time difference between the images. This approach identified minima in the \gls{rms}, which correspond to intervals where the coma morphology repeats, which is assumed to be correlated with the nucleus rotation period. The validity of this technique has been demonstrated in previous studies. For instance, \cite{2015A&A...574A..38O} applied a similar methodology to comet C/2012 F6 (Lemmon) and determined a rotation period of $9.52\pm0.50$ hours.

\begin{figure*}
    \centering
    \begin{minipage}[b]{0.46\textwidth}
        \includegraphics[width=\textwidth]{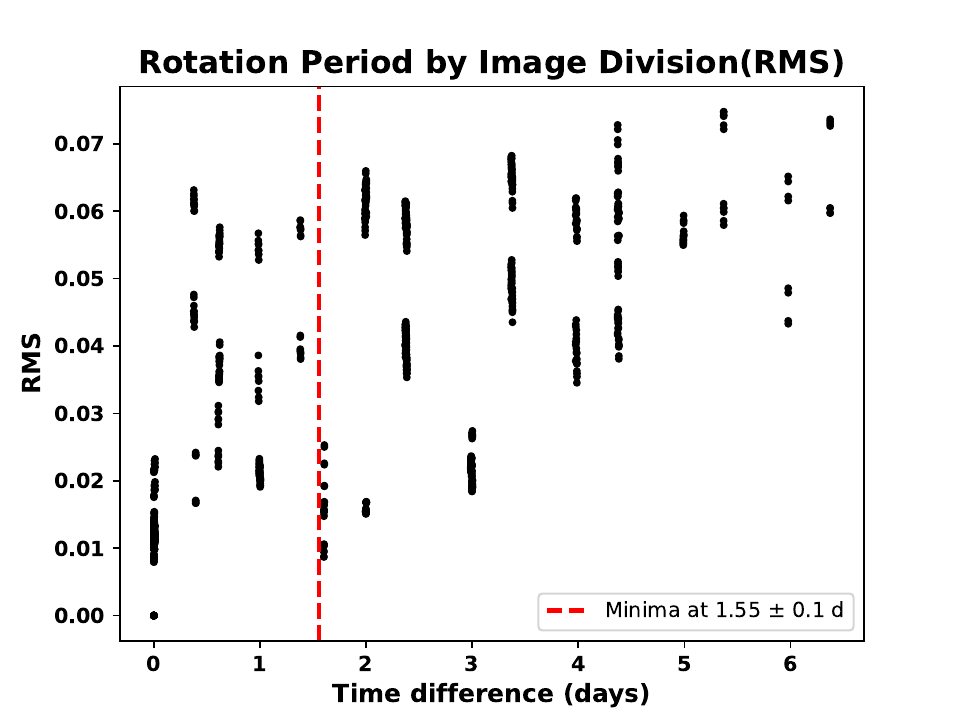}
    \end{minipage}
    \hfill
    \begin{minipage}[b]{0.46\textwidth}
        \includegraphics[width=\textwidth]{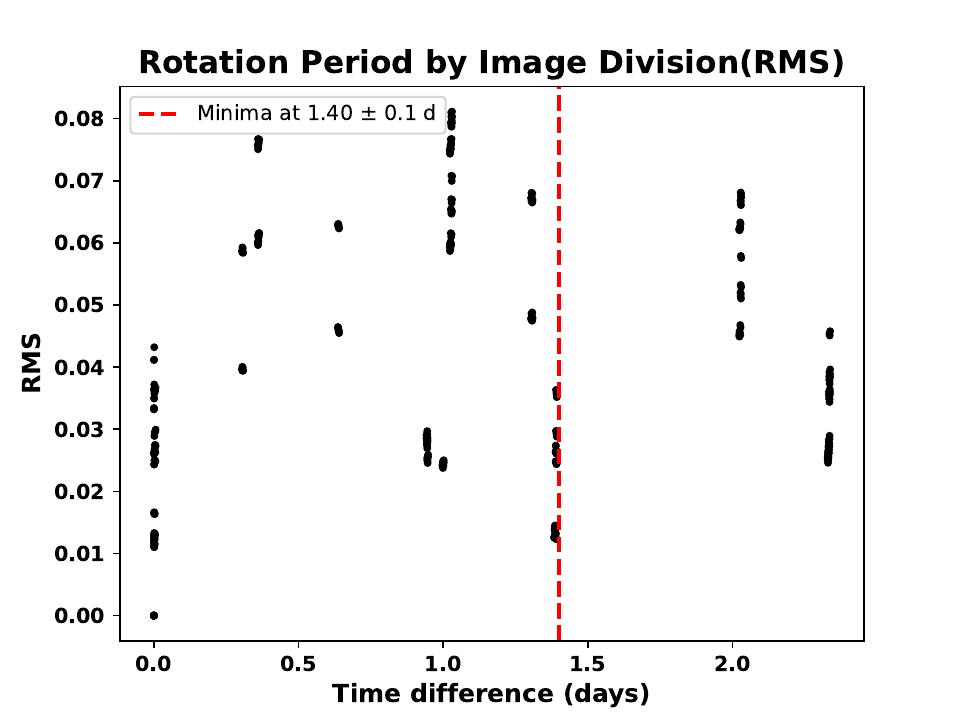}
    \end{minipage}
    \caption{These show \gls{rms} of pairwise-divided, normalised CN images plotted against the time difference between observations from 29 January–4 February (left) and 26–28 February 2007 (right). The \gls{rms} minima mark epochs where two images share nearly identical morphology, suggesting the same or very close rotational phase. A pattern of repeating minima is seen, with principle minima found at 1.55$\pm$0.10 d in the first run and 1.40$\pm$0.10 d in the second (see red the dotted lines). This is because the observational cadence and temporal coverage was such that a clear minima could only be detected after three complete rotation cycles, meaning the measured time difference between minima corresponds to three times the rotation period, yielding rotation periods of 11.8$\pm$0.5 h and 11.3$\pm$0.5 h respectively. }
    \label{fig:Rotation}
\end{figure*}

This method was applied to the two distinct observing campaigns: January 31 to February 4, 2007, and February 25 to 28, 2007. The \gls{rms} minima observed in figure~\ref{fig:Rotation} showing the \gls{rms} verses time difference plots are indicative of the timescale over which the coma morphology repeats due to the rotation of the nucleus. The \gls{rms} of the residuals minimized for pairs separated by multiples of 11.8$\pm$0.5 hours and 11.3$\pm$0.5 in the first and second observing runs respectively, shows recurrence of morphology at this interval(see left and right panels in the figure \ref{fig:Rotation}).

\subsubsection{Rotation Period by Angular Displacement}
As a complementary approach to the \gls{rms} analysis, a second technique was employed to constrain the rotation period of C/2006 P1. This method was developed because the limited temporal coverage in our dataset prevented the \gls{rms} approach from detecting rotation periods shorter than 9.5 hours. By tracking the angular displacement of morphological features in the coma over time, particularly the high-contrast CN jets, we could probe faster rotational modulation that might not be apparent from \gls{rms} minima alone. 

 For each image, the angular position of the primary (brightest jet) was determined through a multi-step procedure. First, the images were re-binned azimuthally in annuli centered on the optocenter to produce jet intensity profiles as a function of \gls{pa}. These profiles were generated approximately 2500 km ($\sim 3$ arcsec) from the nucleus to minimize noise and enhance signal from near-nucleus structures. The profile of each jet was fitted using a one-dimensional Gaussian function. The fitting was also performed over a restricted angular range around the peak of the profile to avoid contamination from other features or background asymmetries. The parameter $\theta_c$ obtained from the fit was adopted as the jet's central \gls{pa} for the corresponding image.

 Assuming constant angular velocity, the rotation rate $\omega$ is given by:

\begin{equation}
    \omega = \frac{\Delta \theta}{\Delta t},
\end{equation}

where $\Delta \theta = \theta_2 - \theta_1$ and $\Delta t = t_2 - t_1$. The rotation period, $P$, was then derived as:

\begin{equation}
    P = \frac{2\pi}{\omega} = \frac{360^\circ}{\omega} = \frac{\Delta t}{\left( \frac{\Delta\theta}{360} \right)}
\end{equation}

To account for uncertainties due to variability in the shape, size and contrast of coma the structures as well as  observational noise and projection effects, this process was repeated for multiple independent features over different observing epochs. Only those structures whose morphology and orientation could be confidently matched across multiple images were included, and especially those taken the same night within the closest observing epoch possible. We do not see a clear angular displacement of the distinct CN jet features in the images taken in a single observing night, with the largest temporal separation being $\pm 0.4$ hours. The overall uncertainty in \gls{pa} measurements was roughly $\pm 5^{\circ}$. This uncertainty arises from factors such as the choice of the range or bin size and location, the placement of the jet cutouts, and the fitting parameters, including the adopted sigma, width, and amplitude used in the Gaussian profile fitting. The lack of measurable change in jet \gls{pa} beyond the method’s uncertainties, indicates either no detectable jet motion within this interval or that the comet is not rotating fast enough for such changes to be resolved. Consequently, while the method alone does not set a strict lower limit, it rules out shorter subharmonics of 11.8 h, the tentative period derived from the \gls{rms} analysis, such as $\sim 5.9$ h  and $\sim 3.9$ h, which would produce detectable angular displacements. 
Comet nuclei with rotation periods shorter than 6 hours are generally considered fast rotators, as they approach or exceed structural stability limits \citep[e.g.][]{1997EM&P...79...35J, 2004come.book..281S}. Only a small number of comets have been reported with rotation periods in this regime, and these are typically linked to unusual physical properties or the action of strong non-gravitational torques. The absence of any significant variation in the jet \glspl{pa} of C/2006 P1 over 0.4 h  therefore indicates that its nucleus is not an extreme fast rotator.

\subsection{Rotational Phase}
To test the rotation period determined in the previous subsections, we arrange enhanced CN-filter images in phase order. We test a series of possible periods around the previously determined best fit. This was done using two approaches as shown in the next subsections, starting with 5.65 h, half the rotation to 35.5 h, three times the period, and around where we find the minima from the \gls{rms} analysis method.

\subsubsection{Rotational Phasing}
\begin{figure*}
    \centering

    \includegraphics[width=\textwidth]{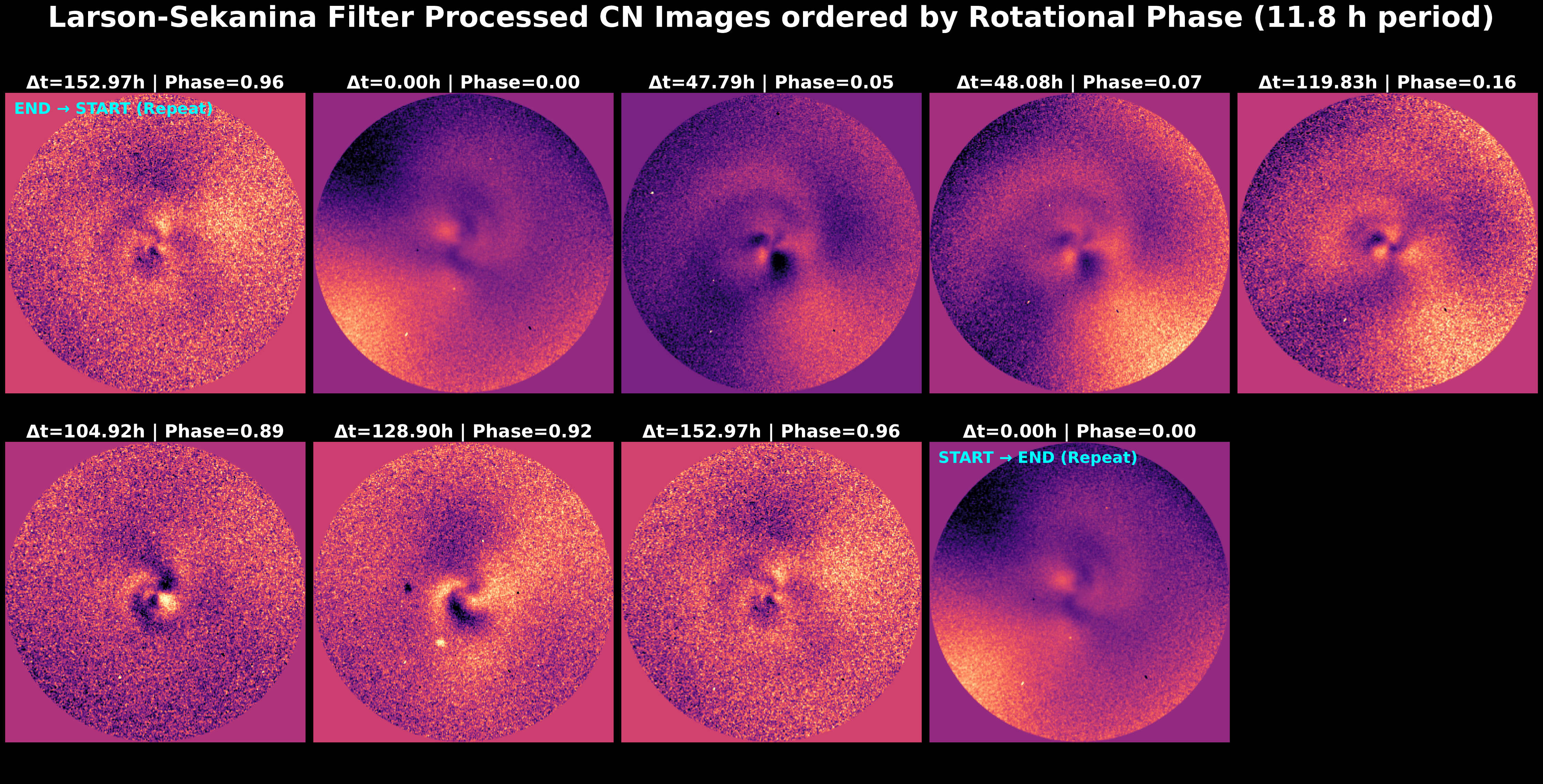}
    
    \vspace{0.4cm}

    \includegraphics[width=\textwidth]{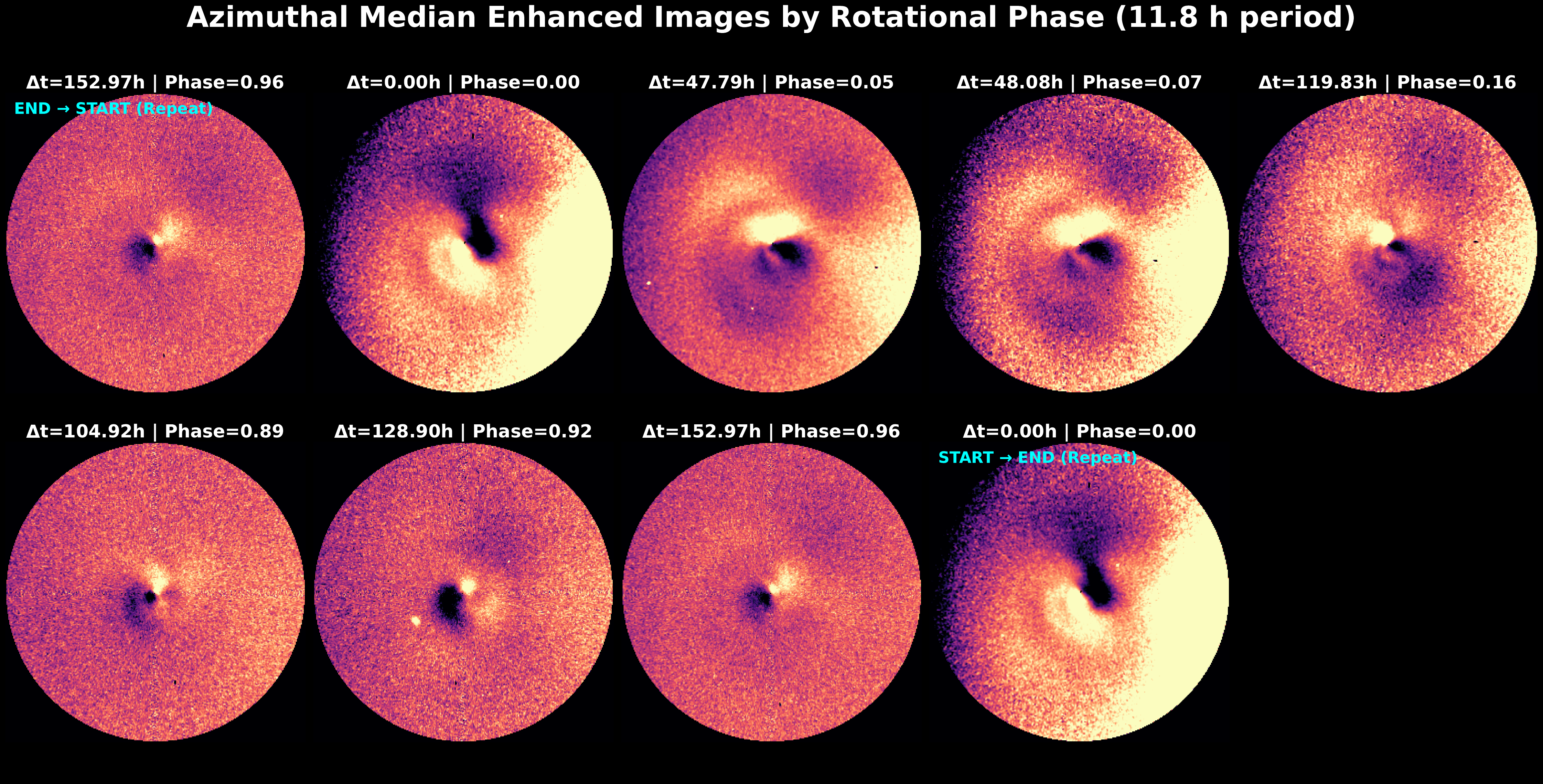}

    \caption{CN images in the first observing run ordered by rotational phase (adopting an 11.8 h rotation period) showing the systematic evolution of inner-coma morphology across the rotation cycle. The top panel shows images processed with Larson-Sekanina filter, revealing narrow jets, spiral and subtle arcs that shift in \gls{pa} with phase. Near phase 0.0, a bright jet with curved structures on the other side dominates; assuming the same jet, by phases 0.05–0.15, it shifts position and develops into curved-like structure with new jets structures emerging.  Near phases 0.89–0.96, a bright collimated jet reappears then develops into a corkscrew-like structure and finally almost identical configuration reappears. The bottom panel shows azimuthally enhanced CN images also arranged by rotational phase (11.8 h period). At phase 0.0, a distinct jet and subtle arcs emerges; between phases 0.05–0.16, the structures persists with a second jet emerging forming nearly symmetrical jets structure with the brightest jet also  becoming sharper; and by phases 0.89–0.96, nearly collimated jet with close orientation, \glspl{pa} are seen. The recurrence of nearly the same large-scale morphology at around phases 0.0 and 1.0 demonstrates periodic behaviour and supports the 11.8 h rotation period. We also note inconsistent pattern between phases 0.96 and 0.05 i.e, at phase 0.00, on 29 January image, where the jet-\gls{pa} deviate from the pattern. }
    \label{fig:Phase}
\end{figure*}

Rotational phasing was done by assigning each of the selected images a phase, setting zero phase at the first image. Images were then ordered by phase to trace the jet’s evolution. Overally, when the assumed period matched the true rotation rate, the jets shifted smoothly and continuously while incorrect periods produced discontinuities or irregular jumps in jet motion.
In order to implement this method, we assigned a reference phase $\phi = 0$ to the time of the first image on the 29 January 2007 where we could first see a dominant jet, i.e. the brightest. Each frame was then assigned a rotational phase using assumed rotation period starting from 5.65 h, about half the lower limit of the adopted period, and stepped in 0.05 h trial increments to 34.5 h, about thrice the period during period testing, and fine-tuning by steps of 0.001 h near plausible rotation periods, and finally ordered by phase. 
As shown in the upper panels in figure \ref{fig:Phase}, the 11.8-hour folding of images taken in the first observing run and processed by Larson and Sekanina method, a part from inconsistencies seen in phases 0.00 and 0.05, jets exhibit a general smooth behaviour, some features gradually shift and fade, others intensify and then disappear as the nucleus rotates, producing a sequence of appearing and vanishing jets. The lower panels in the same figure \ref{fig:Phase} show representative images enhanced with division by azimuthal median method and phase ordered using the 11.8 h rotation, the jets are seen shifting in \glspl{pa} with rotation though incoherent pattern is observed between phases 0.96 and 0.05 (i.e, the 29 January image at phase 0.00), where the jet-\gls{pa} shift deviate from the otherwise consistent trend. Incorrect trial periods produce clear discontinuities, features jump forward or backward between successive phase steps, which guided acceptance limits for the period. Challenges with this approach included variable image quality with varying S/N ratios, limited and uneven temporal sampling, artifacts from the image processing methods and possible changing viewing geometry and scale of the observed features, all of which can mimic or mask true motion, therefore, we used a considerate approach so that small discontinuities from these effects were tolerated rather than rejected. Other plausible periods, though showing some incoherent patterns are 7.765 h, 10.955 h, 11.025 h, 11.450 h, 11.3 h, 22.6 h.

\subsubsection{Phase Diagram}
\begin{figure*}
	\includegraphics[width=0.8\textwidth]{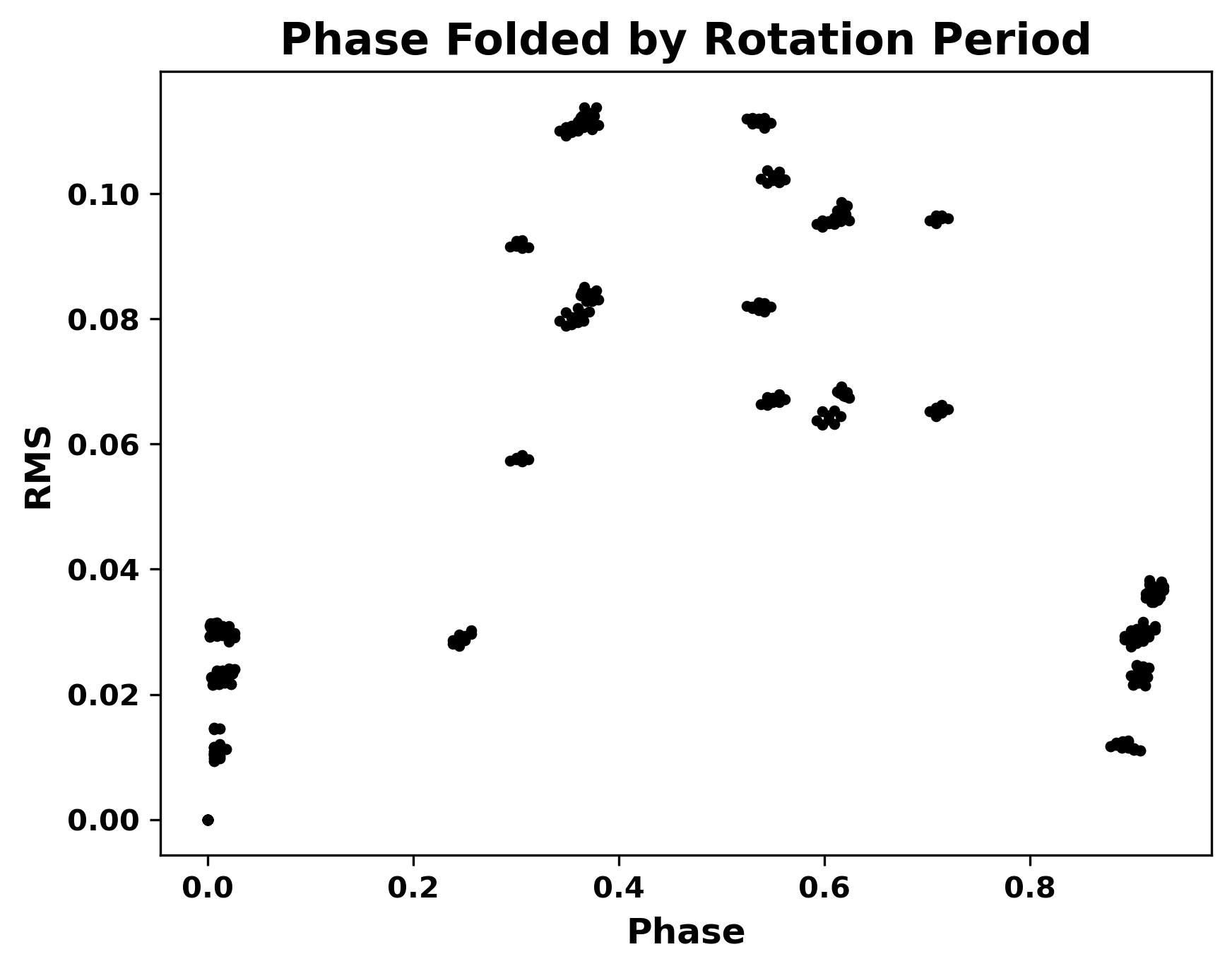}
    \caption{Phase-folded \gls{rms} analysis used to verify the 11.8$\pm0.5$ h rotation period of C/2006 P1. For each trial period, CN images were pairwise divided to generate residual maps, and \gls{rms} values measured in an annulus (3000–7500 km) to quantify morphological differences, then phase-folded. Folding the data with the adopted 11.8$\pm0.7$ h period produces  minima phase 0 and 1, and maxima 0.5, revealing repeatable jet morphology. The agreement of \gls{rms} analysis and \gls{pd} method minima supports the derived rotation period.}
    \label{fig:Phase_folded}
\end{figure*}

To verify the 11.8 h rotation period and assess the temporal repeatability of C/2006 P1's CN jet morphology,  we combined \gls{pd} with the \gls{rms} variability analysis. Similar to the rotational phasing of the images, \gls{pd} was used to look for the rotation period that makes the \gls{rms} repeat in the most consistent way when phase folded into cycles. In order to implement this, each normalized CN images were divided by each other producing residual maps that emphasize on time and phase dependent features quantified by their \gls{rms}. The \gls{rms} of intensity differences within an annular region (3000 – 7500 km cometocentric distance) was then calculated to quantify morphological deviation between phase and time separated pairs. For each assumed period $P$, the phase difference was defined as  

\begin{equation}
\Delta \phi = \left( \frac{t_2 - t_1}{P} \right) \bmod 1,
\end{equation}

and corresponding \gls{rms} values were plotted against $\Delta \phi$ as shown in figure \ref{fig:Phase_folded}. Here, we again started by 5.65 h half the lower limit of the adopted rotation period of 11.8 h $\pm0.5 $ upto twice the period, increasing by 0.05 h. Distinct minima near $\Delta \phi \approx 0.0$ and $1.0$, and maxima near $0.5$, indicated repeatable structural changes consistent with rotation. In contrast, folding the data with an incorrect period produced no coherent phase-dependent pattern, with the data mostly randomly distributed across phase. The alignment of the \gls{rms} minima in the \gls{pd} supports the adopted rotation period of 11.8 h $\pm0.5$, demonstrating that this period reproduces the repeating evolution of the jet morphology.

\section{Discussion}
We estimate a tentative rotation period of about 11.8$\pm0.5$ hours from analysis by the \gls{rms} variability method, but a few inconsistencies in the phase plots and limited observational coverage mean we cannot confirm it. Earlier work by \citet{2010KPCB...26..322K} estimated a rotation period of approximately 21 hours based on tail simulations, but this value remained unverified by direct coma or nucleus observations. In our analysis, we find the 21-h period inconsistent with the behaviour revealed by it's coma morphology and particularly rotational phasing. The 11.8-h period from time-resolved CN morphology here represents the first observational determination of the comet’s rotation and is approximately half the 21-h period inferred from the modeling \citet{2010KPCB...26..322K}. Even if the actual rotation period is uncertain, our comparison of image pairs revealed no measurable angular displacement of distinct jet features over separations shorter than $\sim$24 minutes, indicating no detectable rapid jet motion and thus ruling out the possibility that C/2006 P1 is an extremely fast rotator. The observed recurrence of almost similar CN jet structures at consistent \glspl{pa} in images spaced by integer multiples of the rotation period suggests the presence of relatively stable, anisotropic outgassing patterns from persistent active areas on the nucleus. A slight difference of approximately 0.5 h between the periods independently determined for the two observing runs based on the \gls{rms} technique, separated by $\sim$21 days, may indicate a small temporal change in rotation state, although this deviation remains fully within the uncertainty range of the technique and therefore it is hard to conclude that the difference reflects genuine spin-up behaviour.

Previous studies reveal fast, medium, and long-period comet rotators, with examples ranging from very short periods such as 96P/Machholz ($\sim$4.10 h; \citealt{2019AJ....157..186E}) and 322P/SOHO ($\sim$2.8 h; \citealt{2016ApJ...823L...6K}) to intermediate values like 46P/Wirtanen (8.94–9.14 h; \citealt{2021PSJ.....2....7F}), 73P (10.38 h; \citealt{2019AJ....158..112G}), Hale–Bopp ($\sim$11.3 h; \citealt{1998ApJ...501L.221L}) and C/2001 Q4 (17.6–23.2 h; \citealt{2007AJ....133.2001F, 2004IAUC.8349....1L}), and much longer periods such as 9P/Tempel 1 (41.85 h; \citealt{2005SSRv..117..137B}), 1P/Halley (7.4 d; \citealt{1986Natur.324..646M}), and 41P’s rapid evolution from 24 to 48 h \citep{2019AJ....157..108S}. Relative to this broader distribution \citep{2017MNRAS.471.2974K}, C/2006 P1's rotation period falls toward the medium class but remains well within the range observed for strongly active comets.

This conclusion is further supported by image pair projection analysis, in which recurring morphological configurations, e.g., in figure \ref{fig:CN_morphology} in the second observing run, images between 26 February at 00:18(in panel A), 27 February at 09:39(panel C) and 28 February at 08:20(panel E) revealed a rotational phase match separated by 11.8$\pm$0.5 hours. The close agreement between this result and the period derived from \gls{rms} variability demonstrates the reliability of morphology-based diagnostics for characterizing rotating, jet-producing cometary nuclei.

The rotational characteristics of a comet nucleus can be linked to its physical properties, including shape, internal structure, and activity distribution \citep{2004come.book..359P, 2010A&A...512A..60V}. Relatively fast rotators such as C/2006 P1 can potentially approach the critical limit for centrifugal breakup, especially if they possess high axial ratios or low tensile strength\citep{2013ApJ...775L..10S}. The limiting rotation rate provides a means to infer the bulk density of cometary nuclei \citep{2006MNRAS.373.1590S}. However, C/2006 P1's observed morphology across two observing runs exhibited stable jets with persistent structure and no signs of disintegration, suggesting that the nucleus can withstand the identified spin rate. This also indicates that the observed activity likely originates from discrete, localized sources rather than widespread surface outgassing.

Furthermore, the CN morphology of C/2006 P1 evolved relatively fast from well-defined spiral and arc-like features in the first run to more linear and fan-like jets in the second. This evolution reflects possible changes in the viewing geometry, heliocentric and geocentric distance, and phase angle, which collectively govern the illumination of active regions and the projection of jets onto the plane of the sky. Between the two observing runs, the phase angle changed from $\sim$65$^\circ$ to $\sim$39$^\circ$ and anti-sun's \glspl{pa} from $\sim$179$^\circ$ to $\sim$188$^\circ$( see table \ref{tab:obs}). This change in aspect can alter the projected jet orientation, curvature, and relative brightness on the sky plane. This is sufficient to account for the observed morphological differences and is therefore the dominant effect compared to the changes in heliocentric and geocentric distances. Such transformations are consistent with behavior seen in comets like Hale-Bopp and Lemmon \cite{1998ApJ...501L.221L, 2015A&A...574A..38O}, where the interplay between rotation and solar insolation produces temporally dynamic yet periodically repeating structures.

The observed linear, fan-shaped, and spiral jet structures suggest discrete and possibly latitude-dependent active regions, though without detailed modeling, such interpretations remain indicative rather than definitive. A comprehensive modeling that accounts for viewing geometry, solar illumination, projection effects and gas dynamics is required to more accurately constrain the locations of active sources, the nucleus spin-axis orientation and rotational parameters \citep{1987ESASP.278..315S, 1987ESASP.278..323S, 2010A&A...512A..60V, 1991ASSL..167..769S, 1984AJ.....89.1408S}, remaining a task for future work.

\section{Summary and Conclusions}

Narrowband CN imaging of Comet C/2006 P1 (McNaught) revealed rapidly evolving jet and arc structures that enabled the first direct observational constraint on its nucleus rotation. Temporal analysis of \gls{rms} variability in images enhanced by division by azimuthal median yielded a rotation period of 11.8$\pm$0.5 hours, representing a tentative measure of the comet's rotation period as this was not fully consistent with the phased images. Additionally, $\sim$0.5 h period difference between two runs separated by 21 days may hint at a slight change in rotation period, but it lies within uncertainties, making genuine spin-up hard to confirm. No significant angular displacement of distinct jet features was detected between image pairs separated by up to $\sim$24 minutes, with any apparent shifts remaining within the uncertainty of the angular-tracking technique. This indicates that the nucleus is not a rapid rotator on timescales shorter than this period. The persistence and recurrence of jet morphology over consecutive rotations suggest stable, localized active regions, while their gradual evolution reflects anisotropic outgassing modulated by nucleus spin and changing solar illumination geometry near perihelion. With a derived period of 11.8$\pm$0.5 h, C/2006 P1 belongs to the middle of the observed distribution, comparable to Hale–Bopp ($\sim$11.3 h; \citealt{1998ApJ...501L.221L}) and slower than 322P/SOHO ($\sim$2.8 h; \citealt{2016ApJ...823L...6K}), yet clearly distinct from the long-period rotators such as 1P/Halley ($\sim$2.1–7.4 d) and 9P/Tempel 1 ($\sim$41 h).

\section*{Acknowledgements}

This work is based on observations collected at the European Southern Observatory under programmes 278.C-5045 and 278.C-5051.

\section*{Data Availability}

Raw data is available from the ESO archive.



\bibliographystyle{mnras}
\bibliography{example} 

\bsp	
\label{lastpage}
\end{document}